\documentclass[conference]{IEEEtran}
\IEEEoverridecommandlockouts
\usepackage{cite}
\usepackage{amsmath,amssymb,amsfonts}
\usepackage{algorithmic}
\usepackage{dutchcal}
\usepackage{graphicx}
\usepackage{textcomp}
\usepackage{booktabs}
\usepackage{xcolor}
\def\BibTeX{{\rm B\kern-.05em{\sc i\kern-.025em b}\kern-.08em
    T\kern-.1667em\lower.7ex\hbox{E}\kern-.125emX}}
\begin{document}

\title{Exploring Filterbank Learning for Keyword Spotting\\
\thanks{This work was supported, in part, by the Demant Foundation.}
}

\author{\IEEEauthorblockN{Iv\'an L\'opez-Espejo$^1$, Zheng-Hua Tan$^1$, Jesper Jensen$^{1,2}$}
\IEEEauthorblockA{\textit{$^1$Department of Electronic Systems, Aalborg University, Denmark} \\
\textit{$^2$Oticon A/S, Denmark}\\
\texttt{\{ivl,zt,jje\}@es.aau.dk, jesj@oticon.com}}
}

\maketitle

\begin{abstract}
Despite their great performance over the years, handcrafted speech features are not necessarily optimal for any particular speech application. Consequently, with greater or lesser success, optimal filterbank learning has been studied for different speech processing tasks. In this paper, we fill in a gap by exploring filterbank learning for keyword spotting (KWS). Two approaches are examined: filterbank matrix learning in the power spectral domain and parameter learning of a psychoacoustically-motivated gammachirp filterbank. Filterbank parameters are optimized jointly with a modern deep residual neural network-based KWS back-end. Our experimental results reveal that, in general, there are no statistically significant differences, in terms of KWS accuracy, between using a learned filterbank and handcrafted speech features. Thus, while we conclude that the latter are still a wise choice when using modern KWS back-ends, we also hypothesize that this could be a symptom of information redundancy, which opens up new research possibilities in the field of small-footprint KWS.
\end{abstract}

\begin{IEEEkeywords}
Filterbank learning, keyword spotting, end-to-end, gammachirp filterbank, gammatone filterbank.
\end{IEEEkeywords}

\section{Introduction}
\label{sec:intro}

Handcrafted speech features such as Mel-frequency cepstral coefficients (MFCCs) and log-Mel features are well-established for many speech applications \cite{Gold11}. Those mimic human perception by roughly simulating aspects of the human auditory system and have shown good performance over the years. However, it is evident that these features are not necessarily optimal for any particular speech processing task and it is reasonable to believe that learned features could lead to better performance.

Thanks to the potentials of deep learning and the increasing availability of speech resources, a recent trend is the development of end-to-end deep learning systems where the feature extraction process is optimal according to the task and training criterion, e.g., \cite{Jung18,Hannah18}. In particular, for applications like speaker verification anti-spoofing \cite{Yu17} and audio source separation and audio scene classification \cite{Zhang19}, optimal filterbank learning has shown improvements with respect to using a standard Mel filterbank.

Filterbank learning has also been explored for automatic speech recognition (ASR) purposes \cite{Sainath15,Seki17,Neil18}. In \cite{Sainath15}, Sainath \emph{et al.} train a raw time convolutional layer (i.e., filterbank), initialized with a gammatone filterbank, jointly with a convolutional, long short-term memory deep neural network (DNN) acoustic model. The front-end learned in \cite{Sainath15}, however, is not able to beat the performance of standard log-Mel features in terms of word error rate (WER). While the approach followed in \cite{Neil18} is very similar to that of \cite{Sainath15}, in \cite{Seki17}, Seki \emph{et al.} consider a pseudo-filterbank layer comprised of Gaussian-shaped filters operating in the power spectral domain. The gains, center frequencies and bandwidths of the pseudo-filterbank layer are trained jointly along with the back-end (i.e., DNN) for ASR. The improvements reported in \cite{Seki17,Neil18} are relatively modest and, moreover, it is unclear whether they are statistically significant, as the authors do not provide confidence intervals along with their WER results.

In this work, we explore filterbank learning for keyword spotting (KWS). To the best of our knowledge, \cite{Simon19} is the only (very recently) reported attempt that integrates filterbank learning in KWS. In \cite{Simon19}, a convolutional neural network (CNN), which is trained to perform keyword prediction from the raw speech waveform, integrates parameterized sinc-convolutions acting as a filterbank. Such a front-end, in which only the cut-off frequencies of the filters are trainable along with the back-end parameters, was already proposed in \cite{Ravanelli19}. Unfortunately, the authors of \cite{Simon19} do not carry out a comparison between using parameterized sinc-convolutions and traditional (i.e., handcrafted) speech features, so the possible advantages of employing a learned filterbank in terms of KWS performance remain unclear.

In this paper, we fill this gap by exploring two different filterbank learning approaches for KWS and comparing them with the use of traditional speech features. First, learning the weights of a filterbank matrix in the power spectral domain is examined. Secondly, we study the utilization of a psychoacoustically-motivated filterbank like the gammachirp \cite{Irino99} (which is an extension of the popular gammatone), where different parameters such as the gains, center frequencies and bandwidths of the filterbank are trainable. For both approaches, the learnable filterbank parameters are optimized by backpropagation jointly with a state-of-the-art KWS back-end consisting of a deep residual neural network \cite{Tang18b}.

From our experimental results, our main observation is that, in general, there are no statistically significant differences between the use of a learned filterbank and handcrafted speech features in terms of KWS accuracy, so we state that traditional speech features are still a good choice when employing modern KWS back-ends. Similarly, Robertson \emph{et al.} \cite{Robertson19} recently reported no statistically significant improvements to phone error rate when using either Gabor- or gammatone-based features instead of standard log-Mel features with a modern end-to-end CNN phone recognizer. In \cite{Robertson19}, they point out the difficulty comparing their work with previous work on learned filterbanks where single error rates are presented instead of statistical analyses of the results over repeated trials. The question is therefore whether those single error rates are meaningful or can be explained by a lucky setting of parameters.

The rest of this paper is organized as follows. In Section \ref{sec:learning}, two different approaches for filterbank learning in the context of KWS are presented. The experimental framework is described in Section \ref{sec:framework}. Then, our experimental results are shown and discussed in Section \ref{sec:results}. Finally, Section \ref{sec:conclusion} concludes this work.

\section{Filterbank Learning for Keyword Spotting}
\label{sec:learning}

In this section, we present two different filterbank learning approaches for KWS. Bear in mind that, for both approaches, the trainable filterbank parameters are optimized by backpropagation jointly with the deep residual neural network-based KWS back-end of \cite{Tang18b} (architecture \texttt{res15}).

\subsection{Filterbank Matrix Learning}
\label{ssec:matrix}

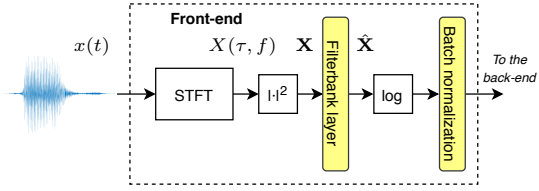
\begin{figure}
\centering
\includegraphics[height=2.45cm]{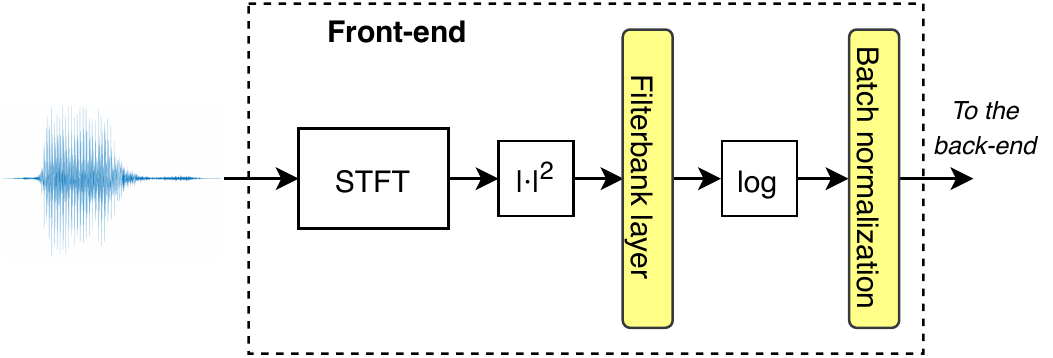}
\put(-178,51){\scriptsize$x(t)$}
\put(-127,51){\scriptsize$X(\tau,f)$}
\put(-94,51){\scriptsize$\mathbf{X}$}
\put(-71,51){\scriptsize$\hat{\mathbf{X}}$}
\caption{Diagram of learnable filterbank matrix scheme.}
\label{fig:matrix}
\end{figure}

Figure \ref{fig:matrix} depicts a diagram of our learnable filterbank matrix scheme. Notice that the front-end diagram is very similar to a log-Mel feature extraction front-end except that the Mel filterbank is replaced by a trainable filterbank.

Let $x(t)$ be a speech signal (possibly containing a keyword) and $X(\tau,f)$ its corresponding short-time Fourier transform (STFT), where $\tau=1,...,T$ and $f=1,...,F$ denote the time frame and linear frequency bin indices, respectively. In addition, $T$ and $F$ refer to the total number of time frames and linear frequency bins, respectively, of the signal. Let
\begin{equation}
 \mathbf{X}=\left[\begin{array}{ccc}
  |X(1,1)|^2 & \ldots & |X(1,F)|^2 \\
  \vdots & \ddots & \vdots \\
  |X(T,1)|^2 & \ldots & |X(T,F)|^2
 \end{array}\right]
\end{equation}
be a $T\times F$ matrix comprised of the squared magnitude of $X(\tau,f)$, then, the filterbank layer applies the following transform to $\mathbf{X}$:
\begin{equation}
 \hat{\mathbf{X}}=\mathbf{X}\cdot h(\mathbf{W}),
\end{equation}
where $\mathbf{W}$ is the learnable $F\times K$ filterbank matrix, $K$ is the total number of filterbank channels and $h(\cdot)$ is an element-wise applied non-linearity to ensure the positivity of the filterbank weights (as similarly considered in, e.g., \cite{Yu17,Zhang19}). In this work, $h(\cdot)=\max(\cdot,0)$ is chosen to be the rectified linear unit (ReLU) function. Then, the result of the logarithmic compression $\log\left(\max\left(\hat{\mathbf{X}},\eta\right)\right)$, where $\log(\cdot)$ and $\max(\cdot)$ are element-wise applied and $\eta=e^{-50}$ is a threshold to avoid numerical issues, is fed to a batch normalization layer the goal of which is to perform feature mean and variance normalization for robustness purposes. Finally, the output from the batch normalization layer is used by the back-end for keyword prediction.

\subsection{Gammachirp Filterbank Learning}
\label{ssec:gammachirp}

\begin{figure}
\centering
\includegraphics[height=2.45cm]{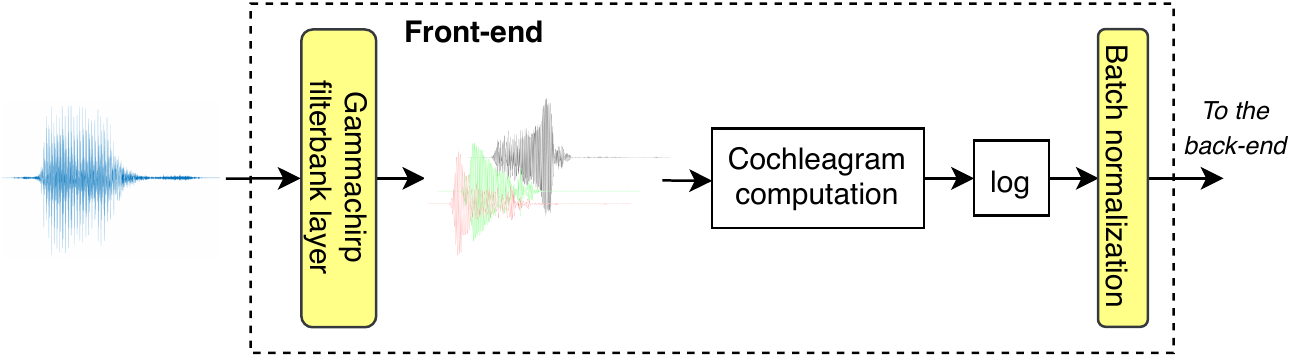}
\put(-225,51){\scriptsize$x(t)$}
\put(-170,51){\scriptsize$x_k(t)$}
\put(-85,51){\scriptsize$\left|\hat{X}(\tau,k)\right|^2$}
\caption{Diagram of learnable gammachirp filterbank scheme.}
\label{fig:gammachirp}
\end{figure}

In this subsection, we consider a psychoacoustically-motivated gammachirp filterbank \cite{Irino99} with learnable parameters. This dynamic auditory filterbank consists of a gammatone filterbank with an additional frequency-modulation term, the so-called chirp term, that yields an asymmetric amplitude spectrum. The chirp term is coherent with physiological observations on frequency-modulations in mechanical responses of the basilar membrane \cite{Irino99}.

The impulse responses of the gammachirp filterbank can be defined as \cite{Irino99}
\begin{multline}
 g_c(t,k)=a_kt^{n-1}e^{-2\pi b\mbox{\scriptsize ERB}(f_k)t} \\
 \times\cos\left(2\pi f_kt+c\log(t)+\phi\right),
 \label{eq:gammachirp}
\end{multline}
where $\{a_k;\;k=1,...,K\}$ are filter gains, $n$ and $b$ define the envelope of the gamma function, $c$ is the chirp term\footnote{Note that if $c=0$, (\ref{eq:gammachirp}) becomes the gammatone filterbank.}, $\phi$ is the initial phase (which is neglected in this work) and $\mbox{ERB}(f_k)$ is the equivalent rectangular bandwidth of the $k$-th filter with center frequency $f_k$. At moderate stimulus levels \cite{Moore90},
\begin{equation}
 \mbox{ERB}(f_k)=24.7+0.108f_k\;\;\; \mbox{[Hz]}.
 \label{eq:erb}
\end{equation}

A diagram of our learnable gammachirp filterbank scheme is outlined in Figure \ref{fig:gammachirp}. The gammachirp filterbank layer implements the linear convolution operation $x_k(t)=x(t)*g_c(t,k)$ ($k=1,...,K$), where $a_k$, $n$, $b$, $c$, $f_k$ and the ERBs are trainable parameters. To preserve the physical meaning of these parameters, the ReLU function is applied to $a_k$, $b$, $f_k$ and the ERBs, whereas $n$ is constrained to be $\max(n,1)$. Then, the cochleagram computation module segments every signal $x_k(t)$ into $T$ overlapping frames of $M$ samples each, $x_{\tau,k}(m)$ ($m=1,...,M$), and estimates the cochleagram $\left|\hat{X}(\tau,k)\right|^2$ by means of Parseval's theorem as
\begin{equation}
 \left|\hat{X}(\tau,k)\right|^2=M\sum_{m=1}^{M}x^2_{\tau,k}(m).
 \label{eq:parseval}
\end{equation}
Finally, logarithmic compression and batch normalization are applied to the cochleagram as discussed in Subsection \ref{ssec:matrix}.

\section{Experimental Framework}
\label{sec:framework}

We use the Google Speech Commands Dataset (GSCD) \cite{Warden18} for KWS experiments. This database consists of 105,829 one-second long speech files with a sampling rate of $f_s=16$ kHz. Each speech file comprises one word among 35 possible candidate words. The GSCD is split into training ($\sim$80\% of the data), validation ($\sim$10\%) and test ($\sim$10\%) sets in such a manner that speakers do not overlap across sets. The deep residual neural network-based KWS back-end of \cite{Tang18b} (architecture \texttt{res15}) is trained to spot the 10 keywords ``yes'', ``no'', ``up'', ``down'', ``left'', ``right'', ``on'', ``off'', ``stop'' and ``go''. Utterances with the remaining 25 words of the GSCD (i.e., non-keywords) are used to define the filler class, so the KWS back-end has to solve an 11-class classification problem. All the word classes are approximately balanced in the different sets.

The length of the analysis window and the hop size are, respectively, $M=480$ and $160$ samples (corresponding to 30 ms and 10 ms at $f_s=16$ kHz). Therefore, every one-second long utterance is comprised of $T=98$ time frames. Furthermore, $F=(M/2)+1=241$ and, as is common \cite{Sainath15,Seki17,Neil18}, $K=40$ is the number of filterbank channels.

The filterbank learning schemes presented in Section \ref{sec:learning} and the KWS back-end are coded by means of Keras \cite{chollet2015Keras}. The back- and front-end are trained by using categorical cross-entropy as the loss function, and Adam \cite{Adam} with default parameters as the optimizer (i.e., the learning rate is 0.001, $\beta_1=0.9$ and $\beta_2=0.999$). Similarly to \cite{Tang18b}, training runs for 26 epochs by default, which is found to be sufficient to guarantee convergence. The size of the minibatch is set to 64 training samples. During training, data augmentation is applied by carefully following the procedure described in \cite{Espejo19}.

As a KWS performance metric, we employ \emph{accuracy}, which is defined as the ratio of the number of correct predictions over the total number of them. To draw meaningful conclusions, accuracy results are provided along with 95\% confidence intervals calculated from outputs of 10 different back-end realizations trained with different random parameter initialization.

\section{Results and Discussion}
\label{sec:results}

\subsection{Filterbank Matrix Learning}
\label{ssec:res_matrix}

We first evaluate our learnable filterbank matrix scheme of Figure \ref{fig:matrix} by jointly and/or alternately training the back- and front-end for a number of epochs. The filterbank matrix of the filterbank layer, $\mathbf{W}$, is initialized by a Mel filterbank. It is worth to note that preliminary experiments explored the initialization of $\mathbf{W}$ by a linear-frequency spaced, triangular-shaped filterbank and no statistically significant differences were observed with respect to the Mel-based initialization.

\begin{table}[t]
  \begin{center}
    \caption{Keyword spotting accuracy results with 95\% confidence intervals, in percentages, from our learnable filterbank matrix scheme.}
    \label{tab:res_matrix}
    \begin{tabular}{l|c}
      \toprule
      \textbf{Test} & \textbf{Accuracy (\%)} \\
      \midrule
      F$\mathtt{f}$B$\mathtt{t}$\_26 (log-Mel) & 95.64 $\pm$ 0.33 \\
      F$\mathtt{t}$B$\mathtt{t}$\_26 & 95.73 $\pm$ 0.24 \\
      F$\mathtt{f}$B$\mathtt{t}$\_26 + F$\mathtt{t}$B$\mathtt{f}$\_10 & 95.73 $\pm$ 0.38 \\
      F$\mathtt{f}$B$\mathtt{t}$\_13 + F$\mathtt{t}$B$\mathtt{t}$\_13 & 95.30 $\pm$ 0.82 \\
      \bottomrule
    \end{tabular}
  \end{center}
\end{table}

\begin{figure}[t]
 \includegraphics[width=\linewidth]{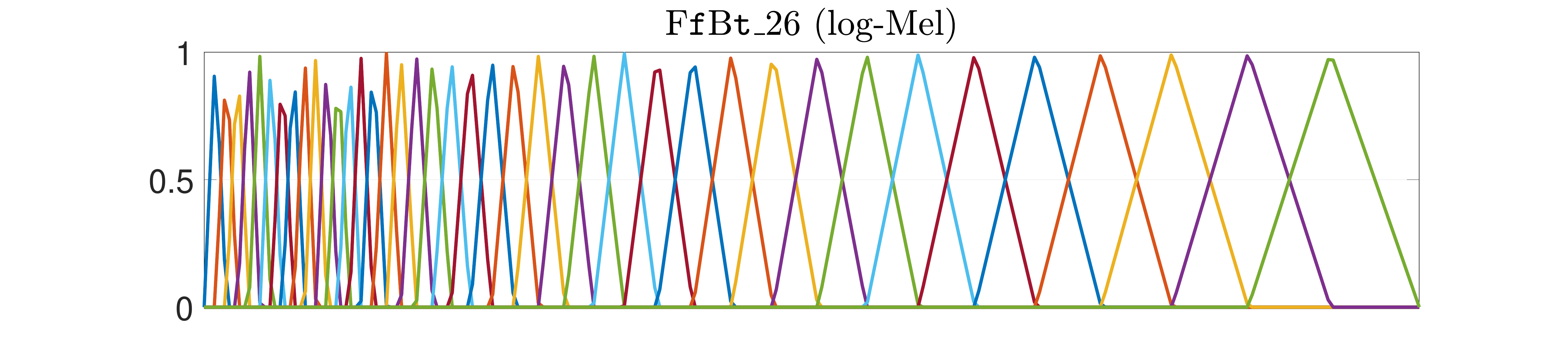} \\
 \includegraphics[width=\linewidth]{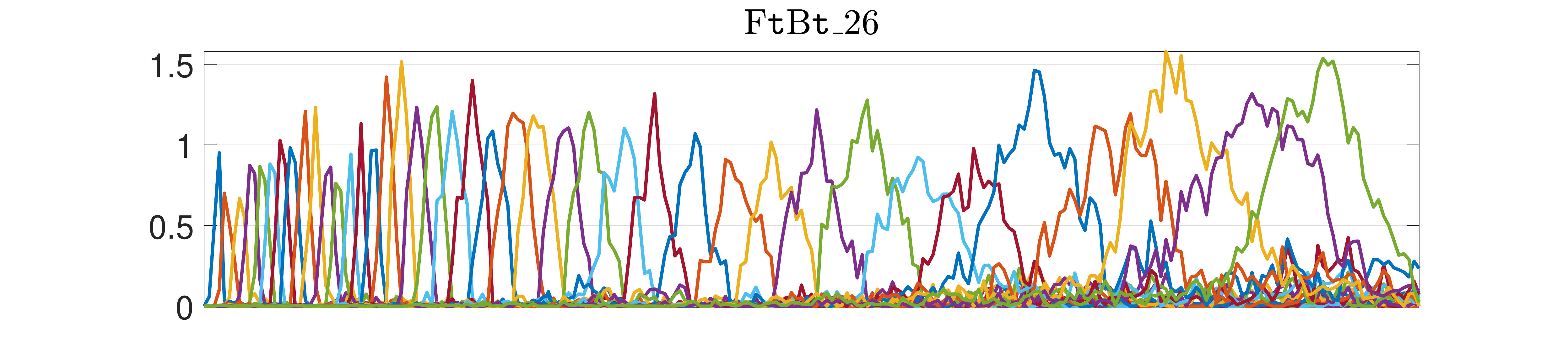} \\
 \includegraphics[width=\linewidth]{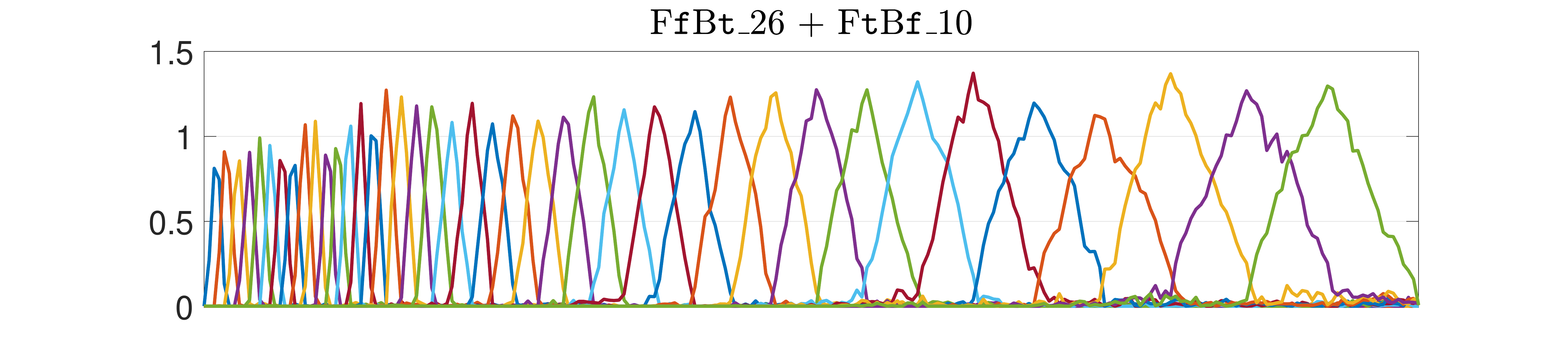} \\
 \includegraphics[width=\linewidth]{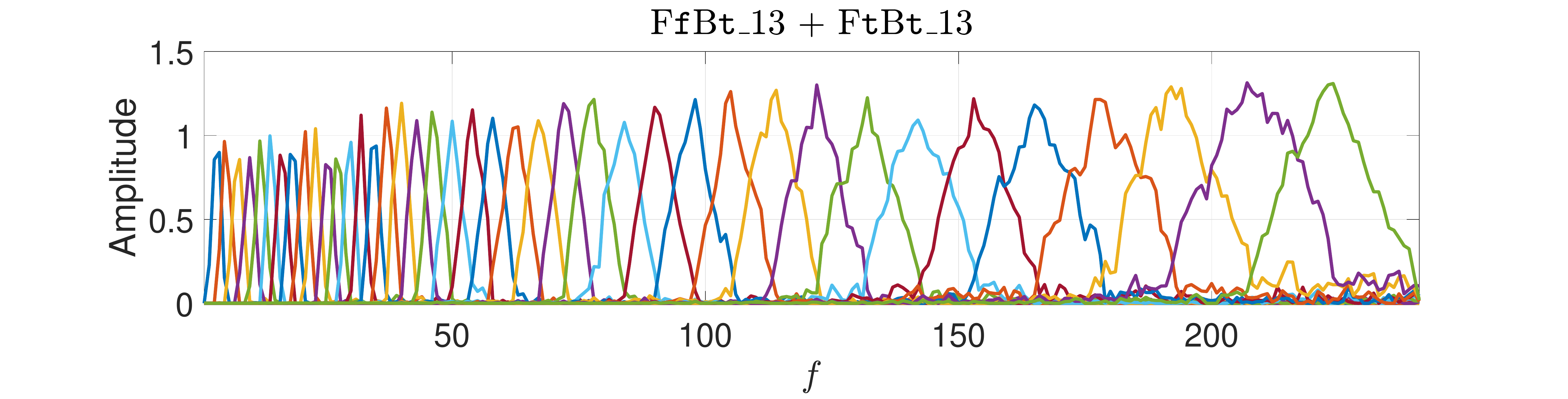}
 \caption{Mel filterbank (top) and average (across the 10 experiment repetitions) learned filterbanks from our learnable filterbank matrix scheme.}
 \label{fig:learned_fb}
\end{figure}

Table \ref{tab:res_matrix} reports our KWS accuracy results from the learnable filterbank matrix scheme by following the naming convention F$\mathsf{x}$B$\mathsf{y}$\_$\mathsf{z}$, where $\mathsf{x}\in\{\mathtt{t},\mathtt{f}\}$ indicates whether the front-end is trained, $\mathtt{t}$, or not (i.e., fixed), $\mathtt{f}$, $\mathsf{y}\in\{\mathtt{t},\mathtt{f}\}$ indicates the same, but for the back-end, and $\mathsf{z}$ is the number of training epochs. Thus, we consider F$\mathtt{f}$B$\mathtt{t}$\_26 a baseline, since it corresponds to the use of standard log-Mel features. As can be seen from Table \ref{tab:res_matrix}, jointly training the back- and front-end (i.e., the filterbank) from scratch, F$\mathtt{t}$B$\mathtt{t}$\_26, does not yield a statistically significant improvement with respect to using standard log-Mel features. Therefore, we assess whether fine-tuning only the filterbank from our well-trained log-Mel baseline by 10 additional epochs, F$\mathtt{f}$B$\mathtt{t}$\_26 + F$\mathtt{t}$B$\mathtt{f}$\_10, provides some performance benefits. According to the results, this choice does not yield a statistically significant improvement, either (95.73\% $\pm$ 0.38 \emph{vs}. 95.64\% $\pm$ 0.33 accuracy). This may be explained by the fact that the back-end is already optimized to work with a Mel filterbank, so substantially altering such a filterbank might even lead to worse performance. This hypothesis is supported by Figure \ref{fig:learned_fb}, which plots the Mel filterbank and learned filterbanks from our learnable filterbank matrix scheme. In this figure, we can see at a glance the relatively higher similarity between the learned filterbank for F$\mathtt{f}$B$\mathtt{t}$\_26 + F$\mathtt{t}$B$\mathtt{f}$\_10 and the Mel filterbank. To no avail, we relax this constraint while still seizing the apparent virtues of the Mel filterbank by training only the back-end from scratch and, prior to convergence, jointly training the back- and front-end, F$\mathtt{f}$B$\mathtt{t}$\_13 + F$\mathtt{t}$B$\mathtt{t}$\_13.

\subsection{Gammachirp Filterbank Learning}
\label{ssec:res_gammachirp}

\begin{table}[t]
  \begin{center}
    \caption{Keyword spotting accuracy results, in percentages, and learned $n$, $b$ and $c$ values from our learnable gammachirp filterbank scheme. Results are provided along with 95\% confidence intervals.}
    \label{tab:res_gammachirp}
    \resizebox{\columnwidth}{!}{\begin{tabular}{l|c||ccc}
      \toprule
      \textbf{Test} & \textbf{Accuracy (\%)} & $n$ & $b$ & $c$ \\
      \midrule
      GT[$\mathtt{f}$]\_I$\mathtt{c}$-Mel & 95.47 $\pm$ 0.36 & 4 & 1.019 & 0 \\
      GC[$\mathtt{f}$]\_I$\mathtt{c}$-Mel & 95.45 $\pm$ 0.58 & 4 & 1.019 & -1 \\
      \midrule
      GC[$\mathtt{t}$]\_I$\mathtt{c}$-Mel & 95.12 $\pm$ 0.42 & 4.69 $\pm$ 0.07 & 0.976 $\pm$ 0.015 & -0.84 $\pm$ 0.05 \\
      GC[$\mathtt{t}$]\_I$\mathtt{c}$-Linear & 95.19 $\pm$ 0.52 & 4.44 $\pm$ 0.05 & 0.866 $\pm$ 0.019 & -0.88 $\pm$ 0.02 \\
      GC[$\mathtt{t}$]\_I$\mathtt{r}$-Mel & 94.68 $\pm$ 0.52 & 4.90 $\pm$ 0.51 & 0.976 $\pm$ 0.115 & -0.97 $\pm$ 0.32 \\
      GC[$\mathtt{t}$]\_I$\mathtt{r}$-Linear & 94.93 $\pm$ 0.45 & 4.65 $\pm$ 0.41 & 0.861 $\pm$ 0.075 & -0.98 $\pm$ 0.38 \\ 
      \bottomrule
    \end{tabular}}
  \end{center}
\end{table}

\begin{figure*}
 \centering
 \includegraphics[width=0.32\linewidth]{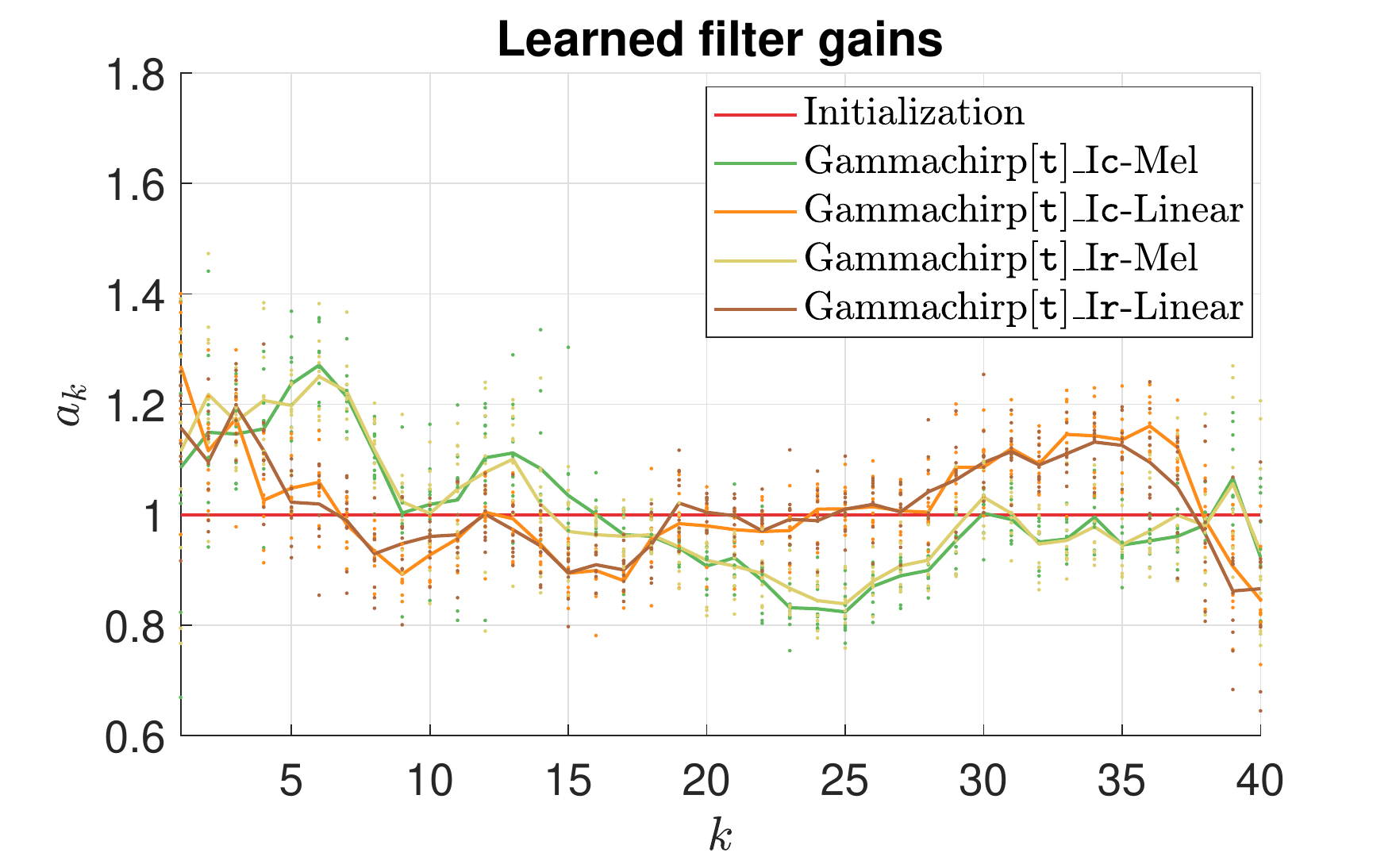}
 \includegraphics[width=0.32\linewidth]{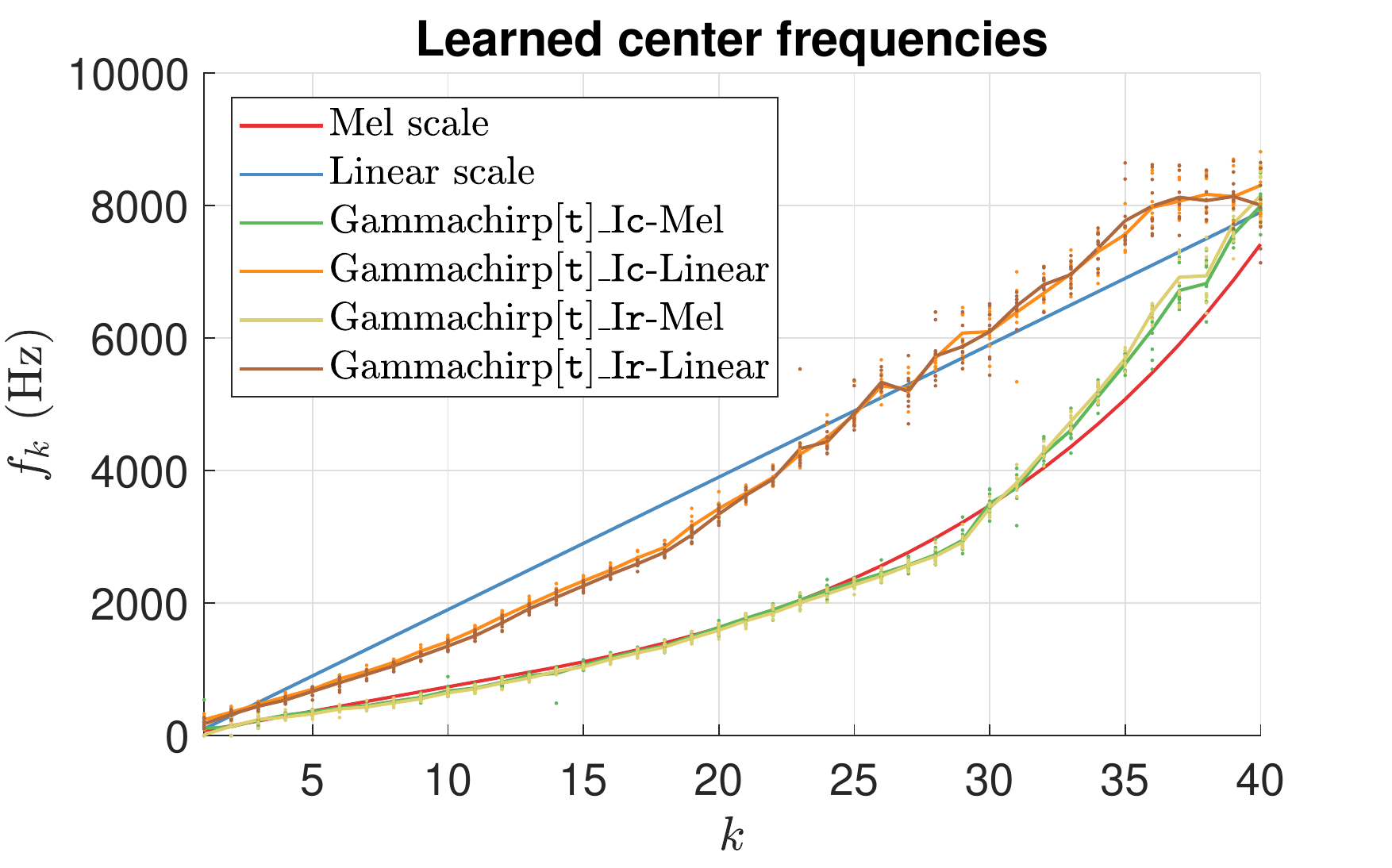}
 \includegraphics[width=0.32\linewidth]{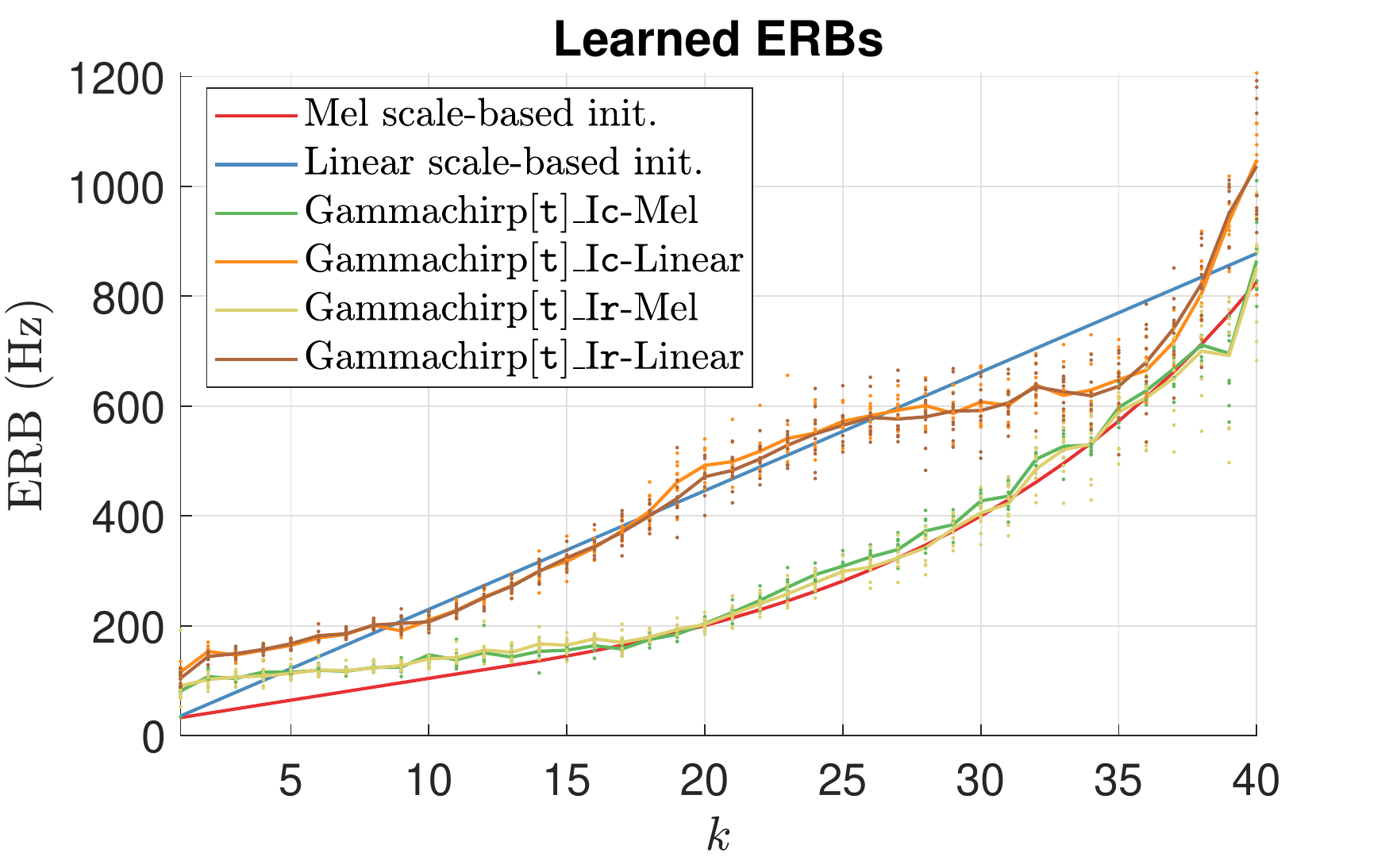}
 \caption{Learned parameter values from our learnable gammachirp filterbank scheme as a function of the filterbank channel $k$. From left to right: filter gains $a_k$, center frequencies $f_k$ and ERBs. Solid lines represent averages across the 10 experiment repetitions (points).}
 \label{fig:learned_params}
\end{figure*}

Table \ref{tab:res_gammachirp} shows our KWS accuracy results and learned $n$, $b$ and $c$ values from our learnable gammachirp filterbank scheme of Figure \ref{fig:gammachirp}. In this case, we follow the naming convention GC[$\mathsf{x}$]\_I$\mathsf{y}$-$\mathsf{z}$, where $\mathsf{x}\in\{\mathtt{t},\mathtt{f}\}$ indicates whether the front-end is trained, $\mathtt{t}$, or not\footnote{In these experiments, the back-end is always trained.}, $\mathtt{f}$, $\mathsf{y}\in\{\mathtt{c},\mathtt{r}\}$ refers to the initialization type of $n$, $b$ and $c$ which can be either constant, $\mathtt{c}$, or random, $\mathtt{r}$, and $\mathsf{z}$ tells whether the center frequencies $f_k$ and the ERBs from (\ref{eq:erb}) are initialized by a Mel or a linear scale\footnote{In \cite{Seki17}, the trained center frequencies of the pseudo-filterbank layer hardly differ from their initialization. As the authors of \cite{Seki17} point out, this can be due to the big difference between the ranges of the center frequencies (i.e., $[0,\;8,\!000]$ Hz) and other DNN weights. We tackle this issue by initializing the center frequencies normalized by $f_s/2$ and de-normalizing them prior to evaluating (\ref{eq:gammachirp}). A similar normalization procedure is carried out for the ERBs.}. When $\mathsf{y}\equiv\mathtt{c}$, the initialization of the gamma function and chirp parameters is $n=4$, $b=1.019$ and $c=-1$ \cite{Irino99}. Otherwise, these parameters are initialized by uniform random sampling according to $n\sim\mathcal{U}(3,5)$, $b\sim\mathcal{U}(0.8,1.2)$ and $c\sim\mathcal{U}(-2,0)$. The impulse responses of (\ref{eq:gammachirp}) are normalized to be in the range $[-1,\;1]$ and $a_k$ is initialized to 1 $\forall k$. Apart from a gammachirp baseline, GC[$\mathtt{f}$]\_I$\mathtt{c}$-Mel, a gammatone baseline, GT[$\mathtt{f}$]\_I$\mathtt{c}$-Mel, is also tested by simply setting $c=0$.

From Table \ref{tab:res_gammachirp}, we can see that there are no statistically significant differences among the different tests in terms of KWS accuracy. Furthermore, standard deviations of the learned $n$, $b$ and $c$ parameters are larger for random initialization than for the constant one. This seems to indicate a certain sensitivity to initial values as well as there are no clear optimal $n$, $b$ and $c$ for the KWS task in terms of accuracy performance. In accordance with Figure \ref{fig:learned_params}, which shows the learned filter gains, center frequencies and ERBs from our learnable gammachirp filterbank scheme, this consideration is equally valid for these parameters, since Mel scale-based initialization leads to rather different learned parameters than the linear scale-based one.

In \cite{Sainath15}, max-pooling is employed for cochleagram derivation instead of (\ref{eq:parseval}). In this equation, notice that $x_{\tau,k}(m)$ results from segmentation of $x_k(t)$ by using a rectangular window. The authors of \cite{Neil18} claim that using a Hann window and the Parseval's theorem for cochleagram computation is superior to using max-pooling in the context of ASR. We have also tried these two approaches and no statistically significant differences were observed with respect to the approach reported in this paper.

Moreover, in \cite{Sainath15}, the learned front-end is unable to beat log-Mel features in terms of WER. The authors of \cite{Sainath15} hypothesize that this can be due to the use of a strong back-end (i.e., acoustic model), though they finally find that this is not a reason when testing on lighter back-ends. Similarly, we explored the utilization of different lighter back-end models (e.g., $\texttt{res8-narrow}$ \cite{Tang18b}) and we observed the same KWS accuracy trends as the ones from using the stronger \texttt{res15}.

Finally, it is important to highlight that, unsuccessfully, we also tried to directly learn the impulse response samples as in \cite{Sainath15,Neil18}.

\subsection{Feature Fusion}
\label{ssec:res_fusion}

\begin{table}[t]
  \begin{center}
    \caption{Keyword spotting accuracy results with 95\% confidence intervals, in percentages, from fusing log-Mel and learnable gammachirp features and reference tests.}
    \label{tab:res_fusion}
    \begin{tabular}{l|c}
      \toprule
      \textbf{Test} & \textbf{Accuracy (\%)} \\
      \midrule
      F$\mathtt{f}$B$\mathtt{t}$\_26 (log-Mel) & 95.64 $\pm$ 0.33 \\
      GC[$\mathtt{t}$]\_I$\mathtt{c}$-Linear & 95.19 $\pm$ 0.52 \\
      \midrule
      Fusion & 95.65 $\pm$ 0.43 \\
      \bottomrule
    \end{tabular}
  \end{center}
\end{table}

Sainath \emph{et al.} \cite{Sainath15} achieve to beat log-Mel features only by fusing the learnable front-end features with them. They argue that this is because of the complementarity of the learned and Mel filterbanks. As before, it is unclear if the reported improvement is statistically significant.

Table \ref{tab:res_fusion} presents the KWS accuracy result from fusing log-Mel features and GC[$\mathtt{t}$]\_I$\mathtt{c}$-Linear, as the linear scale-based initialization may help provide useful complementary information. As we can see, the fusion result is virtually identical to that from employing log-Mel features only, so we might conclude that the learned gammachirp filterbank conveys no additional information for KWS. Other fusion combinations lead to the same conclusion.

\subsection{Filter Removal}
\label{ssec:res_removal}

\begin{figure}[t]
 \centering
 \includegraphics[width=\linewidth]{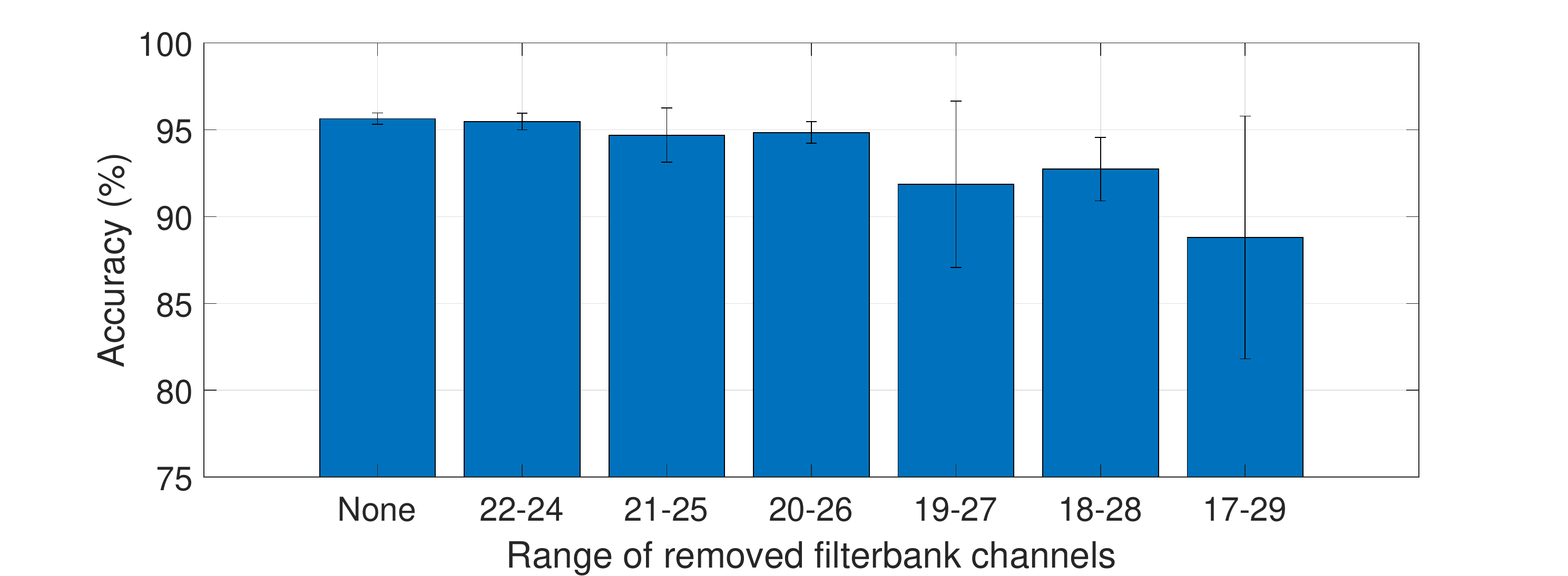}
 \caption{Keyword spotting accuracy with 95\% confidence intervals, in percentages, as a function of the range of removed filterbank channels for the test F$\mathtt{f}$B$\mathtt{t}$\_26 (log-Mel).}
 \label{fig:removed_filters}
\end{figure}

Bearing in mind all of these results, a question emerges: is the filterbank and, in general, the speech feature design actually a crucial part of modern KWS systems? To study this question, we conduct KWS experiments using log-Mel features where we systematically remove filters from the filterbank in order to limit the amount of information available for keyword classification. Filterbank channel removal is carried out around channel $k=23$, the center frequency of which is $f_{k=23}\approx 2,\!000$ Hz, since the frequency band contributing the most to human intelligibility is centered near 2,000 Hz \cite{Paolis96}. Figure \ref{fig:removed_filters} plots KWS accuracy as a function of the range of removed filterbank channels. As can be seen from this figure, performance is negligibly affected even when removing the channels in the range $[20,\;26]$ that spans, approximately, the frequency range from 1,626 Hz to 2,564 Hz. This result supports the hypothesis that KWS systems are fed with a great amount of redundant information. Consequently, this gives clues on why the performance of learned filterbanks and traditional speech features is comparable.



\section{Conclusion}
\label{sec:conclusion}

In this paper, we have explored two different filterbank learning approaches for keyword spotting. Multiple experiments have shown that, in general, there are no statistically significant differences in terms of KWS accuracy between using a learned filterbank and handcrafted speech features, so we conclude that the latter are still a good choice when employing modern KWS back-ends. Furthermore, we have noticed that this could be a symptom of information redundancy, which opens up new research possibilities in the field of small-footprint (that is, low memory and computational complexity) KWS such as the design of much more compact speech features.

\bibliographystyle{IEEEtran}
\bibliography{mybib}

\end{document}